# Formation of nano-size Cr layers on LiNbO$_3$ crystal surfacesby dissimilar electric charges

V. Haiduchok[1], M. Vakiv[1], I.V. Kityk[2], D.Yu. Sugak[3], G. Lakshminarayana[4,*]

[1]*Scientific Research Company "Electron – Carat", 202 Stryjska Str., 79031 Lviv, Ukraine*
[2]*Institute of Optoelectronics and Measuring Systems, Faculty of Electrical Engineering, Czestochowa University of Technology, 17 Armii Krajowej Str., 42-200 Czestochowa, Poland*
[3]*Lviv Polytechnic National University, 12 Bandera, Lviv 79646, Ukraine*
[4]*Wireless and Photonic Networks Research Centre, Faculty of Engineering, Universiti Putra Malaysia,43400 Serdang, Selangor, Malaysia*

## Abstract

Morphology of metal thin layers deposited on polished surfaces of LiNbO$_3$ crystal with dissimilar electric charges has been studied using various methods of vacuum evaporation. The principal goal of the work is to optimizemethods of thin metal coating deposition applied to polar surfaces of LiNbO$_3$crystal. The conducted research has demonstrated that Cr deposition on LiNbO$_3$ crystal faces is characterized by dissimilar polarity results in growth difference conditions of Cr films. The principal role here is played by the sign of the electric charge on the crystal surface along with the type (thermal or magnetron) and conditions observed during sputtering (deposition rate, substrate temperature etc).

***Keywords:***Lithium niobate crystal; thin-film deposition; condensation; electrically-charged surface; film morphology

*Corresponding author– E-mail: **glnphysics@gmail.com; gandham_n@upm.edu.my**

## 1.Introduction

Recently, one can observe an enhanced interest to LiNbO$_3$ crystals due to their possible use in different optoelectronic devices like electro- and acousto-optical modulators, optical delaying lines, triggers, filters, and polarizers etc [1]. Despite a significant efforts on deposition of thin metal films on the surfaces of lithium niobate LiNbO$_3$ single crystals using vacuum deposition methods,it is still very hot topic for the fabrication of acoustic, electronic, optoelectronic, and acousto-optical devices, which employ this single crystal as active element for transforming the external electrical and acoustical in the propagated optical beams [1, 2].It is necessary to emphasize that the electrooptical parameters of the LiNbO$_3$ are very sensitive to intrinsic Li/Nb non-stoichiometry [3,4], which can change the electrooptical efficiencies by more than 30 %. These parameters play a principal role in the processes of transport of ions from the deep bulk states to the surfaces states references.

Therefore, the study of the structure and growth features of such coatings presents a significant practical interest. In addition, LiNbO$_3$is ferroelectric, and simultaneously it demonstratespromising pyroelectric and piezoelectric properties [2]. Heating and/or deformation of lithium niobate single crystal leads to accumulation of dissimilarnon-equilibrium electric charges on their surfaces, which faces that are perpendicular to crystallographic axis Z. For metal film deposition have been used traditional methods of thermal and rf-magnetron sputtering deposition in order to improve their adhesion to the crystal surface, when it undergoes heating treatment. As a result, LiNbO$_3$ crystal faces that are perpendicular to the polarization vector accumulate dissimilar electric charges. Processes of metal atoms' deposition and thin-film coating generation are expected to have different kinetics depending on the type of electric charge on the crystal surface during the deposition process.

Some effects exerted by an electric field on the metal films' growth processes are described in Ref. [5]. Effects of electron beam treatment applied to the target surface of film deposition during the deposition process have been studied in Refs. [6,7]. It was established that in case of a substrate undergoing electron bombardment as deposition proceeds, a higher concentration of nucleation centers along with better structure and higher film growth homogeneity are observed.However, the results put forth in these papers are not entirely applicable to cases of metal films being deposited on electrically charged surfaces of a ferroelectric crystal such as lithium niobate, since the charge of film deposition target surfaces can be positive or negative alike. Therefore, the main goal of this study is an exploration of the surface of metal layers deposited on lithium niobate crystal faces possessing dissimilar electric charges.

## 2. Experimental

Metal films have been applied using methods of thermal and magnetron deposition with combined facility made by TORR International (USA) on the surfaces of 6×8×0.8 mm wafers sliced off LiNbO$_3$ crystal with the wafer planes being perpendicular to crystallographic axis Z. Crystallographic alignment of faces, their sawing, filing, and polishing was performed along with polarity determination before deposition.

The thickness of metal films and their deposition rate were controlled with SQC-330 quartz sensor with accuracy up to 0,1 nm. The heating temperature of lithium niobate single crystal wafers, whose surfaces metal films were deposited on, was selected within the temperature range 150–200ºC. Obtained films were at least 2 nm thick and their surface homogeneity was equal to 15–20 nm. Film deposition rate was equal to 0.5 Å/s.

Morphology of films deposited on lithium niobate wafer faces of dissimilar polarity was studied using Solver P4 atomic-force microscope and Akashi DS 130C scanning electron microscope.

Optical properties of films were studied with Shimadzu UV-3600 spectrophotometer (Japan) with spectral resolution 1 nm the spectral resolution range 200–2500nm.

## 3. Results and discussion

### 3.1. Film growth processes

Cr film in Cr-Cu system is deposited directly onto the crystal surface. Its primary goal is to improve an adhesion of all the subsequent system layers to material's surface. Since Cr film has significantly lower conductivity as compared to Cu and it should be of a minimal thickness, while simultaneously being of highest quality. In order to solve this task, the effect of technological modes of Cr sputtering during film deposition on the dissimilar polarity faces of lithium niobate single crystals was studied.

The commencement of film growth on crystal surface is determined by the energy barrier of nucleation, i.e. by the formation of a critical nucleus with the high positive free energy required for its further expansion, which would ultimately result in an intact coating. Presence of electric charges on the substrate surface leads to an alteration in nucleus formation energy [8], which in turn affects the whole film generation process. Under conditions of high nucleation barrier, the deposited film will consist of a small number of large-scale nuclei, whereas a film generated at law nucleation barrier will consist of a large number of rather small-scale nuclei because of small minimum stable nucleus size and higher nucleation rate. Such film would consist of tightly fit small islands touching one another and fusing together – it would appear homogenous even at the early stages of deposition.

### 3.2. Film morphology

Evaporation that is deposited on the substrate surface during film deposition in vacuum invariably features different fractions of various impurities regardless of the sputtering method used.

As Cr evaporates, condensing vapors include $Cr^+$ ions along with its compositions $CrO^+$, $CrN^+$ [8]. Having found a negatively charged face of the $LiNbO_3$ crystal, positive ions firmly attach themselves to its surface, becoming the centers of nucleation. Consequently, homogenous metal films without visible islands form on a negatively charged surface of LiNbO$_3$ at a thickness as low as ~ 2 nm (Fig.1).

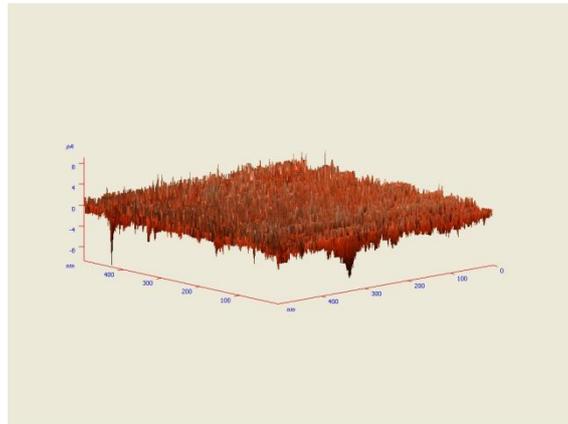

Figure 1. 2 nm thick Cr film deposited using the method of thermal evaporation on the negatively charged surface of LiNbO$_3$ crystal (image from Solver P4 atomic-force microscope)

Chrome film deposited on positively charged lithium niobate crystal faces had island structure at 2 nm thickness (Fig.2). The space resolution and quality of the islands should be improved.

This result is evidently related to the repulsion of positively charged chrome ions and its compositions are characterized by relatively low kinetic energy from lithium niobate surface, leading to higher nucleation barrier.

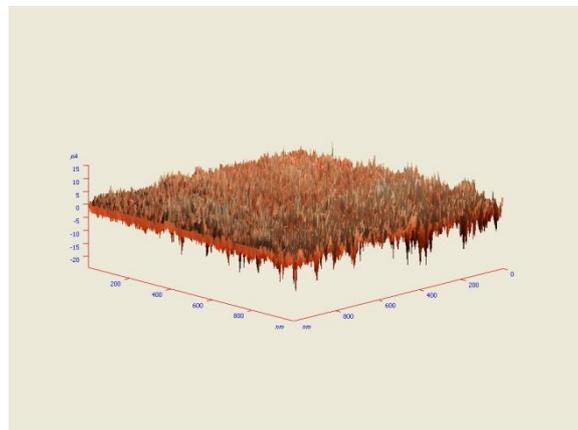

Figure 2. 2 nm thick Cr film deposited using the method of thermal evaporation on the positively charged surface of LiNbO$_3$ crystal (image from Solver P4 atomic-force microscope)

With respect to the thermal method, positive ion concentration in vapors and their kinetic energy show a significant rise during magnetron deposition due to the method's character – the bombardment of chrome target located within the electric and magnetic fields' impact area with Ar$^+$(argon) ions. This deposition method has allowed us to achieve relative homogeneity of chrome films deposited on negatively and positively charged LiNbO$_3$ crystal faces alike, even for a layer 2 nm thick (Figs. 3, 4), with an only slight discrepancy in the film structure.

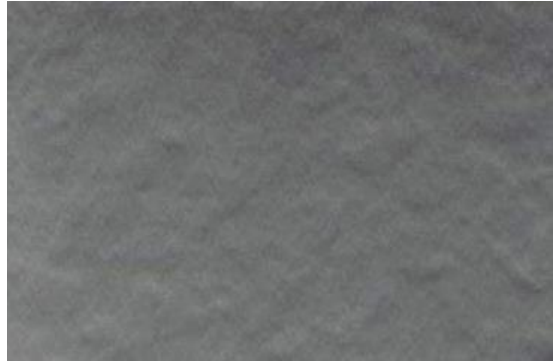

Figure 3. 2 nm thick Cr film deposited using the method of rf-magnetron sputtering in the vacuum on negatively charged LiNbO$_3$ crystal surface (from Akashi DS 130C scanning electron microscope, magnified 5000-fold)

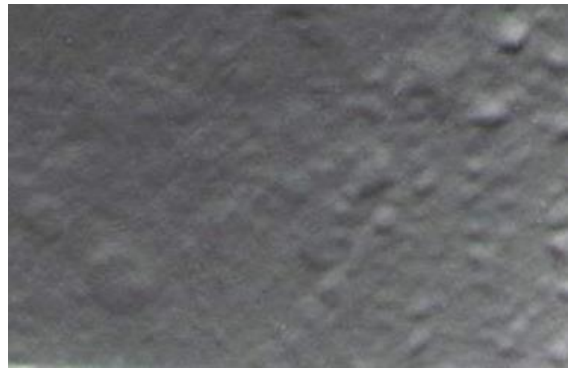

Figure 4. 2 nm thick Cr film deposited using the method of rf-magnetron sputtering in the vacuum on a positively charged LiNbO$_3$ crystal surface (from Akashi DS 130C scanning electron microscope, magnified 5000-fold)

An increase in sputtered particles' kinetic energy during magnetron deposition of films results in lower nucleation barrier.Morphology studies of the films deposited on lithium niobate crystal faces with positive potential have shown that increasing temperature of lithium niobate crystal with fixed chrome layer, deposition rate remains the same leading to a larger size of critical nuclei with island structure persisting. This may be explained by the rise in the condensed particles' mobility on the base surface brought about by its temperature elevation, which in turn makes their interaction and subsequent fusing more likely.

It has also been discovered that the tendency towards island structure aggregation strengthens if film deposition rate is lowered.The process of chrome interlayer formation on polar faces of lithium niobate crystal directly determines the quality of a subsequently deposited copper film.

*3.3. Optical properties of theobtained films*

The difference in the structure of chrome thin films deposited by magnetron sputtering on the surface of LiNbO$_3$ wafers with dissimilar electric potential can be further proved by transmittance spectra of 2 nm thick chrome films within the wavelength range between 300 and 2600 nm (Fig. 5).

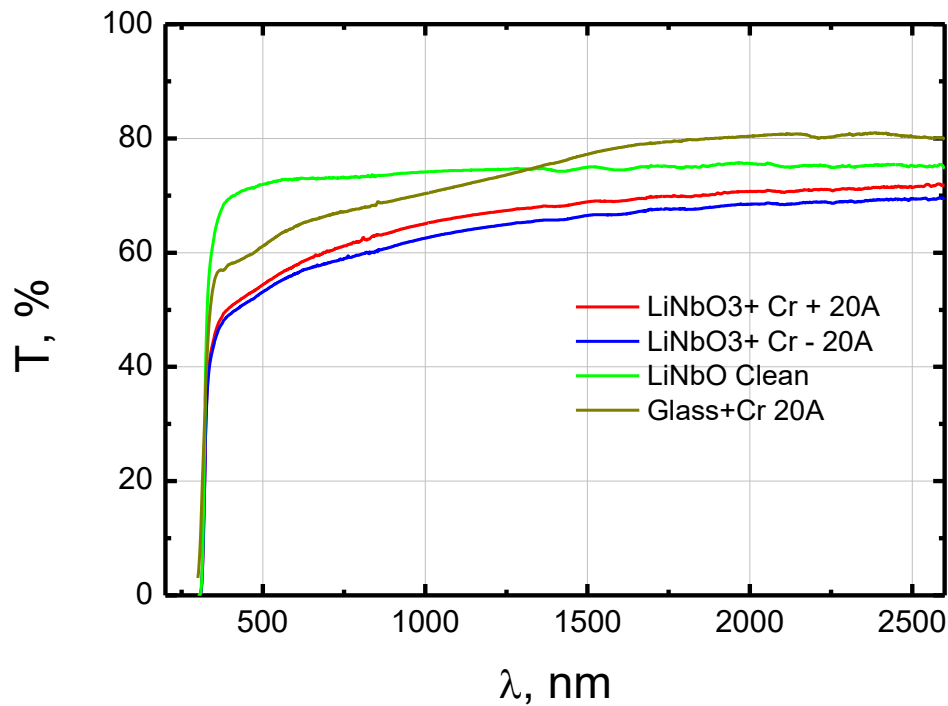

Figure 5. Transmittance spectrum of pure a LiNbO$_3$ wafer (green), 2 nm thick Cr film deposited on a LiNbO$_3$ wafer on the positively charged face (black) and negatively charged face (red)

The spectra distinctly demonstrate a slight but obvious difference in parameters depending on the charge sign of the surface that serves asthe basis for deposition. The lower transmittance of the chrome film deposited of the negatively charged face of LiNbO$_3$ crystal wafer bears testimony to tighter and more homogeneous arrangement of critical nuclei (islands) on this face as compared to the similar film deposited on the positively charged face of the wafer. It should be noted that the films in question were applied within the same deposition cycle in identical conditions.

To understand the observed dependences of transmittance we have performed bands structure and molecular dynamics simulation of the Cr modified layers. The calculationswere done using the band structure band on norm-conserving pseudopotential with procedure similar to the described in the Refs[9,10]. We have considered LNB clusters of different effective sizes. The detailed algorithms for the corresponding matrix elements are presented in the Ref. [11].

We have used the plane wave basis set with a cutoff up to 370 Ry to achieve the eigenvalues reproducibilityup to 0.02 eV. Solving the secular equation was performed using the Querry limited method with the adding of the 50–80 additional plane waves in the Lowdin perturbation approach to carry out the calculations in a more extended plane wave basis set. Electron screening effects were calculated using the Ceperley–Alder expression parameterized by Perdew–Zunger in a form described in the Ref. [10]. The special *k*-point method of Chadhi–Cohen was applied to calculate the electron charge density space distribution. The latter is used to form a charge densityscreening functional for electrons.

Acceleration of the self-consistent procedure was realizedby mixing the (*m*-1)th iteration with 60% of the output $\rho$ before their substitution into the next equationWe assumed an accuracy better than $\varepsilon = 0.13\%$ between the input and output iterations to achieve a self-consistency.

Unfortunately most of DFT calculations give underestimated value of energy gap, however, we can obtain the main tendency. The effective LNB cluster was used with varying effective radius of this cluster. So, we have applied an approach with effective potential as a superposition of t long-range crystal contribution and the clusters with appropriate weighting factors. Additional moleculardynamics on the borders was done similarly to the described in the Ref. [9].

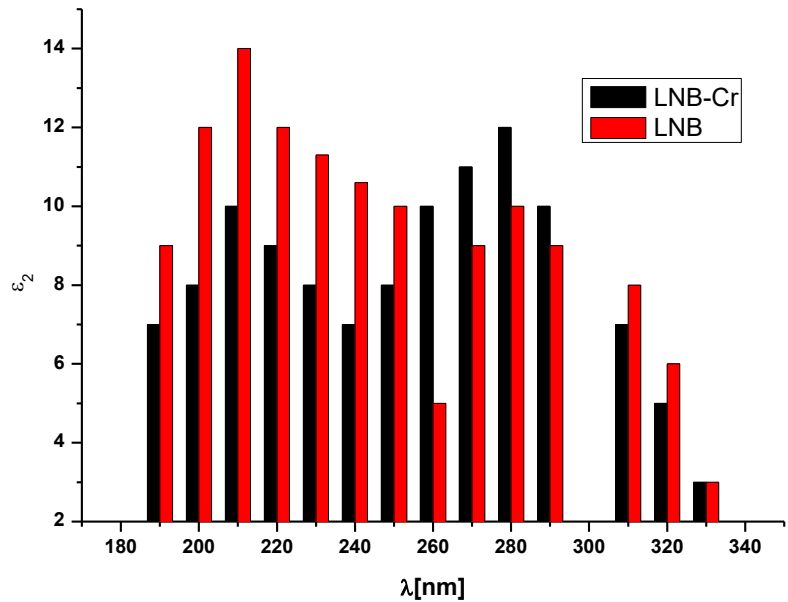

Fig. 6. Theoretically simulated dispersion of the imaginary part of dielectric susceptibility for Theoretically calculated dispersions of dielectric susceptibility $\varepsilon_2(E)$ (see Fig. 6) clearly show that for the Cr doped clusters it occurs a drastic increase of the dielectric susceptibilities with the 260 nm…290 nm. So, this effect is completely determined by the occurrence of the principally clusters, which are very effective for manipulation by optical parameters.

## 4. Conclusions

The conducted studies have shown that when Cr is deposited on $LiNbO_3$, crystal faces have different polarity. The Cr films' growth at the initial stage primarily depends on the charge sign on the crystal surface, which served as the basis for deposition, and on the type (thermal, magnetron) and conditions of sputtering. This one allows, in turn, to optimize the technology of applying metal interlayers to obtain high-quality homogeneous (more parameters) coating with good adhesion regardless of the polarity of target deposition faces. The obtained results can be used for the development of optical and electric devices based on ferroelectric crystals that require a metalized coating to be applied to polar faces in particular. The modeling of this process was confirmed by independent band structure and molecular dynamic simulations.

## Acknowledgments

Authors thank the Condensed system laboratory of Lviv State University and personally

B.Ya. Kulik(Ph.D) for conducting the study of films with atomic-force microscopy. Additionally, this work is a part of the project that has received funding from the European Union's Horizon 2020 research and innovation program under the Marie Skłodowska-Curie grant agreement No. 778156.